\documentclass[prx,reprint,amsmath,amssymb,aps,superscriptaddress,nofootinbib,letter]{revtex4-2}

\usepackage{graphicx}
\usepackage{dcolumn}
\usepackage{bm}
\usepackage{bbm}
\usepackage{appendix}
\usepackage{comment}
\usepackage{url}
\usepackage{soul}
\usepackage[english]{babel}

\usepackage{subfigure}  

\usepackage[colorlinks, linkcolor=black,anchorcolor=blue,citecolor=blue,urlcolor=blue]{hyperref}

\usepackage{siunitx}
\usepackage{dsfont}
\usepackage{enumitem}

\newcommand{\bra}[1]{\left\langle#1\right|}
\newcommand{\ket}[1]{\left|#1\right\rangle}
\newcommand{\braket}[2]{\left\langle #1 | #2 \right\rangle}

\newcommand{\matel}[3]{\left\langle #1 \right| #2 \left| #3 \right\rangle}

\begin{document}

\preprint{APS/123-QED}


\title{Distributing entanglement at the quantum speed limit in Rydberg chains}

\author{Kent Ueno}
\affiliation{Department of Physics and Astronomy, University of Waterloo, Waterloo, Ontario N2L 3G1, Canada.}
\affiliation{Institute for Quantum Computing, University of Waterloo, Waterloo, Ontario N2L 3G1, Canada.}

\author{Alexandre~Cooper}
\email[]{alexandre.cooper@uwaterloo.ca}
\affiliation{Department of Physics and Astronomy, University of Waterloo, Waterloo, Ontario N2L 3G1, Canada.}
\affiliation{Institute for Quantum Computing, University of Waterloo, Waterloo, Ontario N2L 3G1, Canada.}

\date{\today}

\begin{abstract}
We numerically study the transport of Rydberg excitations in chains of neutral atoms. We realize an effective flip-flop interaction using off-resonant driving fields. By tuning the relative distances between atoms and applying atom-selective detuning fields, we realize the perfect transport condition. This condition enables the transfer of a single Rydberg excitation from one end of the chain to the other, allowing the distribution of entanglement across the chain at the quantum speed limit. Through numerical simulations, we identify the set of control parameters that maximizes the transport probability for experimentally relevant parameters. We study the various competing trade-offs involved in the hierarchy of approximations used to map the native Rydberg spin model onto the effective model driving spin transport. Our results suggest that entanglement can be distributed over chains of more than fifty atoms spanning hundreds of microns at room temperature. This study informs the selection of parameters for the experimental realization of perfect transport in Rydberg chains, providing a new approach to distribute entanglement among distant atoms in quantum processors.
\end{abstract}


\maketitle

\section{Introduction}\label{sec:intro}
The efficient distribution of quantum information between distant quantum registers is a fundamental challenge in the development of scalable quantum information processors~\cite{NielsenChuang2000}. A typical approach relies on quantum channels capable of transmitting quantum information with high fidelity~\cite{Schumacher1996}. In quantum processors with static architectures, where registers are fixed in space and their interactions decay according to a power law, linear chains of spins have been proposed as a quantum channel for enabling short-distance quantum communication~\cite{Bose2003, Yao2011}. By precisely tuning the interaction strengths to achieve the perfect transport condition, the quantum state of a spin can be perfectly transferred to another spin located at a mirror-symmetric location in the chain at the quantum speed limit~\cite{Christandl2004, Christandl2005, Yung2006}.
This coherent state transfer enables faster information transport than sequential gate-based protocols and serves as a primitive for entanglement distribution between quantum registers, which in turn enables quantum teleportation protocols~\cite{Bennett1993}.

\begin{figure}[t!]
\includegraphics[]{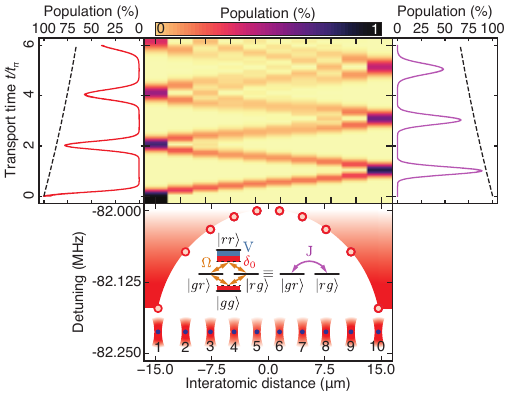}
\caption{
\textbf{Perfect transport in a Rydberg chain.}
(Top)~A single-spin excitation prepared at one end of the chain is coherently transported to the other end at time $t_\pi$. The probability of measuring the single-spin excitation on the first (left) and last (right) spins undergoes collapses and revivals, which are damped by radiative decay (dashed line) and the loss of coherence caused by approximation errors and long-range interactions.
(Bottom)~The interatomic distances and local detunings are tuned to realize the perfect transport condition.
(Inset)~The effective spin exchange between two atoms is realized by detuning the driving field away from the single-spin resonance. The spin exchange is mediated through two-photon Raman transitions between the zero-excitation manifold and the two-excitation manifold, whose energy is shifted due to the Rydberg blockade. The large detuning suppresses the injection of spin excitations into high-excitation manifolds.
}
\label{fig:transport}
\end{figure}


While the theoretical framework for perfect state transfer is well established, its experimental realization has remained so far elusive~\cite{Ciaramicoli2007, Cappellaro2007}, because of limitations imposed by decoherence, disorder, and the difficulty of precisely engineering interactions in solid-state quantum systems. Recent advances in the control of large, interacting quantum many-body systems offer an imminent opportunity for practical realization~\cite{Blatt2012, Preiss2015, Browaeys2020, Karamlou2022, Ungar2024}. A particularly promising platform is offered by configurations of neutral atoms excited to a Rydberg state, also known as Rydberg atom arrays~\cite{Urban2009, Bernien2017, Barredo2018, Ebadi2021}. 
By realizing quantum spin models with variable parameters on arbitrary geometries with more than a thousand of atoms, these quantum simulators act as analog twins of solid-state spin systems, enabling the validation and cross-benchmarking of various control protocols across different parameter regimes.

In this paper, we investigate the conditions required to achieve perfect transport in chains of neutral atoms with Rydberg interactions. Our goal is to identify the set of control parameters that maximizes the probability of perfect transport under realistic experimental conditions, as well as to understand how this probability scales with physical parameters such as the principal quantum number, temperature, and chain length. Through extensive numerical simulations, we identify and quantify the key trade-offs involved in optimizing transport probability, including the competition between transport speed and spin-excitation injection, as well as the interplay between longer lifetimes, stronger interactions, and reduced driving rates at higher principal quantum numbers. Our approach relies on mapping the transverse-field Ising spin Hamiltonian realized by Rydberg chains onto an effective Heisenberg~XX Hamiltonian~\cite{Yang2019, Kim2024}, and engineering the perfect transport condition by tuning the interatomic spacing and applying atom-selective detuning fields.

Beyond finding the set of parameters that maximizes the transport probability, we aim to set a baseline for the experimental study of spin transport on Rydberg platforms and to quantify their robustness against experimental imperfections. For example, the transport probability may be used as a metric to certify the realization of Heisenberg models in Rydberg systems. We also seek to identify the key challenges involved in relying on spin transport to distribute entanglement between distant registers in quantum processors.

The outline of the paper is as follows. We first review the conditions required to realize perfect transport in Rydberg chains and discuss the hierarchy of approximations involved in mapping the theoretical model onto the experimentally-relevant Rydberg model (see Sec.~\ref{sec:models}).
We then present our numerical simulation results identifying the set of control parameters that maximizes the transport probability under realistic experimental conditions (see Sec.~\ref{sec:tradeoffs}). These results highlight some of the trade-offs involved when optimizing over the detuning and principal quantum number. We finally explore the maximum chain length over which the system can serve as a coherent quantum channel for entanglement distribution (see Sec.~\ref{sec:entanglement}). These findings lay the groundwork for implementing quantum state transfer protocols in programmable Rydberg platforms.

\section{Perfect transport in Rydberg chains}\label{sec:models}

\begin{figure}[t!]
\includegraphics[width=8.6cm]{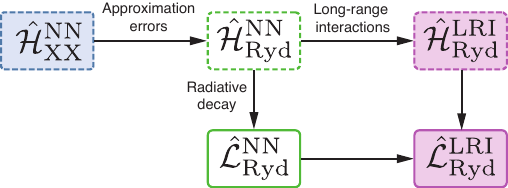}
\caption{
\textbf{Hierarchy of approximations.}
The perfect transport condition is realized for the Heisenberg~XX Hamiltonian with nearest-neighbor (NN) interactions in the absence of radiative decay (blue). Its mapping onto the experimentally-relevant Rydberg model incurs errors due to approximation errors (green), long-range interactions (LRI, purple), and radiative decay.
}
\label{fig:diagram}
\end{figure}

We consider a quantum system formed by a finite chain of $L$ rubidium-87 atoms with variable, possibly non-uniform, interatomic spacings. Each atom encodes a qubit between its ground state, $\ket{g}=\ket{5S_{1/2}}$, and excited Rydberg state, $\ket{r}=\ket{nS_{1/2}}$. The principal quantum number is taken to be $50 \leq n \leq 80$. Each qubit is coherently driven via a two-photon Raman transition mediated through an intermediate electronic state, $\ket{e}$. The detuning of the optical field driving the $\ket{g}$ to $\ket{e}$ transition, $\delta_{ge}$, is chosen to be sufficiently large to allow for the adiabatic elimination of $\ket{e}$ in the dynamics. For simplicity, we refer to the dressed states of the effective two-level qubit as $\ket{g}$ and $\ket{r}$. The effective Rabi frequency between $\ket{g}$ and $\ket{r}$ is $\Omega = (\Omega_{ge} \Omega_{er}) / (2 \delta_{ge})$, where $\Omega_{ge}$ and $\Omega_{er}$ are the Rabi frequencies for the transitions from $\ket{g}$ to $\ket{e}$ and from $\ket{e}$ to $\ket{r}$, respectively. The residual admixture of $\ket{e}$ in $\ket{r}$ is $|\epsilon_r|^2\approx \Omega_{er}^2/4\delta_{ge}^2$, so that the effective decay rate of $\ket{r}$ is $\Gamma_{r}^\prime(n,\theta) \approx \Gamma_{r}(n,\theta) + |\epsilon_r|^2 \Gamma_{e}$, where $\Gamma_{e}$ is the decay rate of the intermediate excited state and $\Gamma_{r}(n, \theta)=\Gamma_{r}^0(n)+\Gamma_{r}^\text{BBR}(n,\theta)$ is the state-dependent decay rate of the bare Rydberg state, which depends on the ambient temperature, $\theta$, due to coupling to the black-body radiation (BBR)~\cite{Beterov2009}. Similarly, the decay rate of the dressed ground state $\ket{g}$ is $\Gamma_{g}\approx|\epsilon_g|^2\Gamma_e$ with $|\epsilon_g|^2\approx \Omega_{ge}^2/4\delta_{ge}^2$; however, we neglect this effective decay rate in our calculations.



The Hamiltonian describing the dynamics of the system is analogous to the transverse-field Ising model~\cite{Bernien2017},
\begin{eqnarray}\label{eq:HRyd}
\hat{\mathcal{H}}_\text{Ryd} = \sum_{l=1}^L \left(\frac{\Omega_l}{2} \hat{\sigma}_x^l - (\delta_0 + \delta_l) \hat{n}_l \right) + \sum_{l<l'} V_{ll'} \hat{n}_l \hat{n}_{l'},
\end{eqnarray}
where $\hat{\sigma}^l_\mu$ is the $\mu$-th Pauli matrix for the $l$-th spin, $\hat{n}_l=\ket{r_l}\bra{r_l}=(\hat{\mathbbm{1}}-\hat{\sigma}_z^l)/2$ is the projection operator onto $\ket{r_l}$, and $V_{ll’}= V(n, \Delta x_{ll'})= C_6(n)/\Delta x_{ll'}^6>0$ is the repulsive van der Waals interaction strength between two Rydberg atoms located at $x_{l}$ and $x_{l'}$ with interatomic spacing
$\Delta x_{ll'}=x_{l}-x_{l'}$. We define the global detuning as $\delta_0=\delta_{er}$ and include an additional site-selective detuning, $\delta_l$, arising from second-order light shifts induced by far-off-resonant, atom-selective laser beams. For the sake of completeness, we choose the Rabi frequency $\Omega_l$ to be tunable on each site, although our protocol does not require it. 




The native spin Hamiltonian does not drive spin transport when the state of the chain is initialized in an eigenstate of the $\bigotimes_j \hat{\sigma}_z^j$ operator~\cite{DiFranco2008a,DiFranco2008}. However, an effective flip-flop interaction can be realized through second-order perturbation theory~\cite{Yang2019, Kim2024}. In the limit of large detunings, $\Omega_l \ll \left|\delta_0\right|, \left|\delta_0 + V_{ll'}\right|$ and $\delta_l \ll \delta_0$, Eq.~\eqref{eq:HRyd} maps onto the effective XX~Hamiltonian~\cite{Yang2019},
\begin{eqnarray}\label{eq:HXX}
\hat{\mathcal{H}}_{\text{XX}} = \sum_{l=1}^L \mu_l \hat{n}_l  + \sum_{l<l'} J_{ll'} (\hat{\sigma}_{+}^{l}\hat{\sigma}_{-}^{l'} + \hat{\sigma}_{-}^{l}\hat{\sigma}_{+}^{l'}),
\end{eqnarray}
where $\hat{\sigma}_\pm^l=(\hat{\sigma}_x^l \pm i \hat{\sigma}_y^l)/2$ are the raising and lowering operators for the $l$-th spin, $\mu_l$ is the effective on-site potential, and $J_{ll'}$ is the effective flip-flop rate. This Hamiltonian is the Heisenberg~XX model in a transverse field, which generates spin transport while conserving the number of spin excitations ($U(1)$ symmetry). 
An analytical expression for the parameters of the effective model, $\{\mu_l, J_{ll’}\}$, expressed in terms of the set of control parameters, $\{x_l,\Omega_l, \Delta_l=\delta_0+\delta_l\}$, can be obtained from second-order perturbation theory~\cite{Yang2019},
\begin{eqnarray}
\mu_l &=& -\Delta_l - \frac{\Omega_l^2}{2\Delta_l} + \sum_{l' \neq l}\frac{\Omega_{l'}^2 V_{ll'}}{4\Delta_{l'}\left(\Delta_{l'} - V_{ll'}\right)}\label{eq:mapping_1},\\
J_{ll'} &=& \frac{\Omega_l \Omega_{l'} V_{ll'}}{8\Delta_l \left(\Delta_l - V_{ll'}\right)}+\frac{\Omega_l \Omega_{l'} V_{ll'}}{8\Delta_{l'} \left(\Delta_{l'} - V_{ll'}\right)}.
\label{eq:mapping_2}
\end{eqnarray}

The effective Hamiltonian from Eq.~\eqref{eq:HXX} can be further simplified by restricting the sum over the dominant nearest-neighbor (NN) interactions, 
\begin{eqnarray}\label{eq:HXXNN}
\hat{\mathcal{H}}^{\text{NN}}_{\text{XX}} = \sum_{l=1}^L \mu_l \hat{n}_l  + \sum_{l=1}^{L-1} J_{ll+1} (\hat{\sigma}_{+}^{l}\hat{\sigma}_{-}^{l+1} + \hat{\sigma}_{-}^{l}\hat{\sigma}_{+}^{l+1}).
\end{eqnarray}
This Hamiltonian is isomorphic to the hard-core limit of the Bose-Hubbard (BH) Hamiltonian via the Holstein-Primakoff transformation~\cite{HolsteinPrimakoff1940}, where the spin excitations play the role of hard-core bosons with at most one particle per site.
By tuning the effective on-site potential, $\{\mu_l\}$, it is thus possible to use the Rydberg chain to study the evolution of bosons in a tilted optical lattice, leading to observables such as Bloch oscillations~\cite{Dahan1996, Preiss2015}.
This Hamiltonian also maps onto a system of free fermions using the Jordan-Wigner approximation~\cite{JordanWigner1928}, enabling an analytical treatment of transport dynamics through the dispersion relation for finite-length chains with open boundary conditions~\cite{Lieb1961}.

\subsection{Perfect transport in Heisenberg chains}

The Heisenberg~XX Hamiltonian generates ballistic transport of single-spin excitations on chains. For equidistant chains with uniform couplings, a single-spin excitation prepared at one end of the chain is not perfectly transported to the other end, unless the effective control parameters are tuned to realize the perfect transport condition~\cite{Christandl2004}. Mathematically, the probability of the system evolving from $\ket{l}$ to its mirror-symmetric counterpart $\ket{L+1-l}$ after a time $t$ is given by the transport probability,
\begin{equation}
p(t) = \left|\matel{{L+1-l}}{\hat{U}(t)}{{l}}\right|^2,
\end{equation}
where $\hat{U}(t)=\exp(-i\hat{\mathcal{H}}t/\hbar)$ is the evolution operator generated by $\hat{\mathcal{H}}$ and $\ket{l} = \ket{g_1g_2 \ldots r_l \ldots g_L}$ is the many-body spin state hosting a single-spin excitation on the $l$-th site for $l\geq1$. We denote the many-body ground state containing no excitations as $\ket{0}=\ket{g_1g_2 \ldots g_L}$.

Perfect state transfer is realized at a time $t_\pi$ if there exists a set of control parameters such that $p_\pi=p(t_\pi)=1$. Multiple choices of parameters may lead to perfect transport~\cite{Christandl2004, Christandl2005, Kay2006}, including for systems with power-law decaying interactions~\cite{Yao2011}. These solutions distinguish themselves by their perfect transport time, which is lower-bounded by the quantum speed limit~\cite{Yung2006}. For the Heisenberg~XX model with NN interactions, the linear spectrum solution~\cite{Christandl2004, Christandl2005} is obtained for
\begin{eqnarray}
    \mu_l &=& \mu \label{eq:constraint_1}\\
    J_{ll+1} &=& {2 J_{\text{max}}}\sqrt{l(L-l)}/\bar{L}\label{eq:constraint_2},
\end{eqnarray}
where
\begin{eqnarray}
J_\text{max} &=& \frac{\Omega^2}{4\delta_0} \left(\frac{\delta_0 \Delta x _\text{min}^6}{C_6} - 1 \right)^{-1}
\end{eqnarray}
is the maximum coupling strength between the central atoms separated by $\Delta x_\text{min}$, and $\bar{L}$ is the characteristic length of the chain, which is defined as $\bar{L}=L$ for even chains, and $\bar{L}=\sqrt{L^2 - 1}$ for odd chains. 
This solution can be understood as the necessary condition to achieve coherent population inversion in a collective spin system~\cite{Cook1979, Christandl2004}. It maximizes the quantum speed limit~\cite{Yung2006} with perfect transport time,
\begin{eqnarray} \label{eq:transport_time}
    t_\pi(L) = \frac{\pi \bar{L}}{4 J_{\mathrm{max}}}.
\end{eqnarray}


\subsection{Solving the inversion problem for Rydberg chains}\label{sec:inversion}
In Rydberg chains, finding the set of control parameters that realizes the perfect transport condition for the linear spectral solution involves solving an inversion problem. This problem requires finding the set of control parameters $\{\Omega_l, \Delta_l, V_{ll'}\}$ that satisfy Eqs.~\eqref{eq:mapping_1}-\eqref{eq:mapping_2} under the constraints imposed by Eqs.~\eqref{eq:constraint_1}-\eqref{eq:constraint_2}. However, this system of equations is underdetermined, featuring $2L-1$ equations and $3L-1$ variables, 
thereby admitting multiple solutions.

\begin{figure*}[ht!]
\includegraphics[]{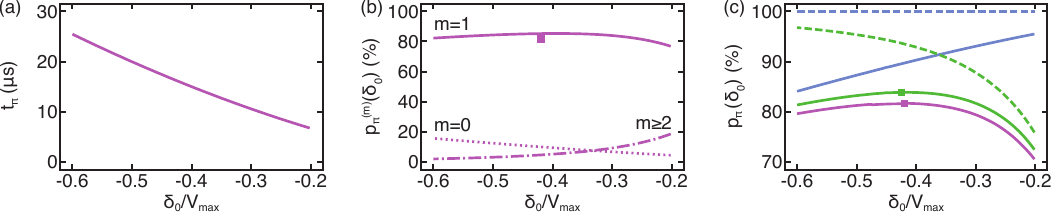}
\caption{
\textbf{Optimizing over the global detuning.}
(a)~The transport time for the Rydberg model, which approximately overlaps for both nearest-neighbor (NN) interactions and long-range interactions (LRI), decreases as the detuning approaches resonance. 
(b)~Fractional populations in the excitation subspaces containing $m = 0$ (dot), $m=1$ (solid), and $m\geq2$ (dash-dot) spin excitations computed for $\hat{\mathcal{L}}^\text{LRI}_\text{Ryd}$ at $t_\pi$. The maximum transport probability (purple square) achieved is less than the total population in the $m=1$ subspace due to imperfect refocusing. 
(c)~Transport probabilities computed in the absence (dashed) and presence (solid) of radiative decay for the Heisenberg~XX model with NN interactions (blue), and the Rydberg model with NN (green) and long-range (purple) interactions. The maximum probabilities (squares) are achieved at distinct optimal values of $\delta_0^*$. All data are computed for $L=16$, $n=70$, and $V_\text{max} = 188.3~\text{MHz}$ for $\Delta x_\text{min}=3~\mu\text{m}$.
}
\label{fig:detuning}
\end{figure*}

A unique solution to the perfect transport problem is obtained for uniformly spaced chains by solving for the restricted set of parameters $\{\Omega_l, \Delta_l\}$~\cite{Yang2019}. Here, we choose to solve for $\{\Delta_l, V_{ll'}\}$ instead, as it is typically easier to adjust the relative positions of neutral atoms using optical traps than to apply atom-selective driving fields. We set the interatomic distance between the central spins to the minimum achievable value, $x_{\text{min}}$, thereby fixing the interaction strength at the center of the chain to $V_\text{max} = C_6(n)/x_{\text{min}}^6$. We also set the on-site detuning of the central spins to zero, $\delta_l = 0$, so that $\Delta_l = \delta_0$. The problem is further simplified by exploiting the mirror symmetry of the perfect transport condition, which imposes $\Delta_l = \Delta_{L+1-l}$ and $x_l = x_{L+1-l}$.

A representative numerical solution for a chain of $L=10$ atoms excited to the $n=70$ Rydberg state is shown in Fig.~\ref{fig:transport}.  
Both the atom-selective detunings $\delta_l$ and the interatomic spacing $\Delta x_{ll'}$ increase away from the center of the chain, ensuring that the on-site potential $\mu_l$ remains constant, while satisfying the linear spectrum condition for the $J_{ll'}$ coefficients. The coherent collapses and revivals of the single-spin excitation at the ends of the chain is damped by radiative decay and the loss of coherence resulting from approximation errors and long-range interactions. We now seek to quantify the competing influence of these effects under realistic experimental conditions.


\section{Optimized control parameters and experimental trade-offs}\label{sec:tradeoffs}

The analytical solution for perfect transport is obtained when the chain evolves under $\hat{\mathcal{H}}^{\text{NN}}_{\text{XX}}$ in the absence of decay, control errors, and other imperfections. In practice, however, the system evolves under the full dissipative Rydberg Liouvillian $\hat{\mathcal{L}}_\text{Ryd}^\text{LRI}$, which includes both radiative decay and long-range interactions (LRI). Realizing the perfect transport condition in a chain of Rydberg atoms requires mapping $\hat{\mathcal{L}}_\text{Ryd}^\text{LRI}$ onto $\hat{\mathcal{H}}^\text{NN}_\text{XX}$. This mapping involves a hierarchy of approximations~(see Fig.~\ref{fig:diagram}), whose validity correlates with the probability of perfect transport, $p_\pi$. Maximizing $p_\pi$ involves a trade-off between competing physical effects, which we quantify using numerical simulations. 

\subsection{Simulation parameters}
\label{sec:simulation_parameters}
We compute $p_\pi$ for different sets of physical and control parameters under realistic experimental conditions. We set the minimum interatomic distance to $\Delta x_{\text{min}} = 3~\mu\text{m}$ to avoid parametric heating caused by the interference of neighboring optical traps. We further set the maximum optical power available to drive the $\ket{e}$ to $\ket{r}$ transition to $P_{er}=4~\text{W}$, which corresponds to a peak intensity of $I = 127~\text{kW}/\text{cm}^2$ when delivered using a Bessel beam with $10~\%$ of its power contained within a core radius of $w_{0}=10~\mu\text{m}$. These constraints effectively set an upper bound for $J_{\text{max}}(n)$, and thus a lower bound for $t_{\pi}\sim J_{\text{max}}^{-1}(n)$.

Fixing the chain length, $L$, and principal quantum number, $n$, we compute the interaction coefficients, $C_6(n)$, and dipole matrix elements, $d_{er}(n)$, using the ARC package~\cite{Sibalic2017}. We use these values to calculate the effective Rabi frequency, $\Omega(n)$, assuming $\Omega_{ge} = \Omega_{er}$ for a given maximum achievable laser power $P_{er}$. Choosing a fixed global detuning value, $\delta_0$, we solve the inversion problem described in Sec.~\ref{sec:inversion} to find the set of local detuning values, $\delta_l$, and atom positions, $x_l$, that realize the perfect transport condition under $\hat{\mathcal{H}}^{\text{NN}}_{\text{XX}}$. Using a reduced subspace method for all chains with greater than $L=4$ spins, we compute the time evolution of the initial state $\ket{1}$ generated by $\hat{\mathcal{H}}^{\text{LRI}}_{\text{Ryd}}$, $\hat{\mathcal{H}}^{\text{NN}}_{\text{Ryd}}$, and $\hat{\mathcal{H}}^{\text{NN}}_{\text{XX}}$ using QuTiP~\cite{Johansson2013}. From the time-evolved state, $\ket{\psi(t)}=\hat{U}(\delta_0,t)\ket{1}$, we compute the transport probability, $p_\pi(\delta_0)=\braket{L}{\psi(t)}$, as well as the population leaked into the many-body ground state containing no Rydberg excitation, $p_\pi^{(0)}=\braket{0}{\psi(t)}$, and the higher-excitation subspaces containing more than one excitation, $p_\pi^{(2^+)}=1-(p_\pi^{(0)}+p_\pi^{(1)})$, where $p_\pi^{(1)}=\sum_{l=1}^{L}\braket{l}{\psi(t)}$. These calculations are repeated for different values of $\delta_0$ to find the optimal detuning, $\delta_0^*=\text{argmax}_{\delta_0}~p_\pi(\delta_0)$, that maximizes the transport probability, $p_\pi^*=p_\pi(\delta_0^*)$. We multiply these values by $\exp(-\Gamma(n, \theta)t_\pi)$ to account for the radiative decay from the Rydberg state. 
This full procedure is then applied for various $n$ and $L$ to quantify the dependence of $p_\pi^*(n, L)$ on the principal quantum number, $n$, and chain length, $L$.

\subsection{Optimizing over the global detuning}\label{sec:detuning}

The first trade-off involves a competition between the speedup of transport dynamics and the breakdown of second-order perturbation theory when sweeping the global detuning, $\delta_0$. As $\delta_0$ approaches resonance, the effective interaction coefficients increase as $J_\text{max}\sim\delta_0^{2}$, leading to a speedup of transport scaling as $t_\pi\sim J_\text{max}^{-1}\sim\delta_0^{-2}$ (see Fig.~\ref{fig:detuning}a). This faster dynamics leads to a reduction in the probability of decay to the zero-excitation ground state, at the cost of an increase in the number of spin excitations being injected into the chain (see the crossover between $p_\pi^{(0)}$ and $p_\pi^{(2+)}$ in Fig.~\ref{fig:detuning}b). As $\delta_0$ is detuned away from resonance, the transport dynamics slows down, increasing the probability of decay to the zero-excitation ground state. We therefore find that for a given set of parameters, there exists an optimal detuning value, $\delta_0^*$ for which the transport probability is maximized~(Fig.~\ref{fig:detuning}c). 

The effect of these competing factors can be quantified by comparing the transport probability obtained from different models over varying values of the global detuning $\delta_0$~(Fig.~\ref{fig:detuning}c). As the detuning is decreased towards resonance, the breakdown of second-order perturbation theory is observed by the drop in $p_\pi$ when comparing the evolution generated by $\hat{\mathcal{H}}_\text{XX}^\text{NN}$ (dashed blue) and  $\hat{\mathcal{H}}_\text{Ryd}^\text{NN}$ (dashed green).
As the detuning is increased away from resonance, the slowdown of transport dynamics 
causes a loss of probability between the exact model generated by $\hat{\mathcal{H}}_\text{XX}^\text{NN}$ (dashed blue) and its decaying version generated by $\hat{\mathcal{L}}_\text{XX}^\text{NN}$ (solid blue). 
The competition between radiative decay and the injection of spin excitations results in a convex curve for $\hat{\mathcal{L}}_\text{Ryd}^\text{NN}$ (solid green) and $\hat{\mathcal{L}}_\text{Ryd}^\text{LRI}$ (solid purple), each of which exhibits a single maximum value at $\delta_0^*$.

\subsection{Optimizing over the principal quantum number}

\begin{figure}[t!]
\includegraphics[width=\columnwidth]{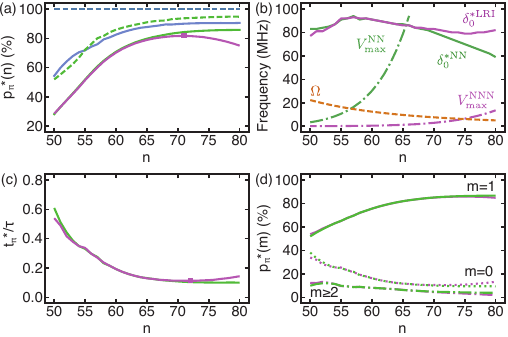}
\caption{
\textbf{Optimizing over the principal quantum number.}
(a)~Maximum transport probability, $p_\pi^*(n)$, achieved for each principal quantum number $n$ in the absence (dashed) and presence (solid) of radiative decay for the Heisenberg~XX model (blue) and the Rydberg model with nearest-neighbor (NN, green) and long-range interactions (LRI, purple). The Rydberg LRI model achieves its maximum transport probability at $n^*=71$ (purple square).
(b)~Maximum interaction strength, $V_\text{max}(n)$, between nearest-neighbor (dash-dot green) and next-nearest-neighbor (dash-dot purple) atoms. These values are compared against the optimal detuning values, $|\delta_0^*(n)| = -\delta_0^*(n)$, for the Rydberg NN (solid green) and LRI (solid purple) models, and the Rabi frequency (dashed orange), which is constrained by the maximum achievable optical power.
(c)~Rescaled transport time $t_\pi^*(n)/\tau(n)$, where $\tau(n)$ is the radiative lifetime of the Rydberg state.
(d)~Fractional populations in the subspaces containing $m = 0$ (dotted), $1$ (solid), and more than $2$ (dash-dot) Rydberg excitations for the Rydberg NN (green) and LRI (purple) models. All data are reported for $L=16$.
}
\label{fig:principal_quantum_number}
\end{figure}

The second trade-off involves a competition between longer lifetimes, stronger interactions, and weaker driving strengths when accessing larger principal quantum numbers. As $n$ increases, the van der Waals interaction strength increases as $V_\text{max}(n)\sim C_6(n)/\Delta x_\text{min}^6 \sim n^{11}$, and the radiative lifetime increases as $\tau(n)\sim n^{3}$, while the effective Rabi frequency decreases as $\Omega(n)\sim d_{er}(n)\sim n^{-3/2}$ when the maximum accessible optical power is fixed. 

We numerically compute the optimal transport probability $p_\pi^*(n)$ obtained for different principal quantum numbers, $n$ (see Fig.~\ref{fig:principal_quantum_number}a). The maximum transport probability monotonically increases with $n$ for both the evolution generated by $\hat{\mathcal{L}}_\text{XX}^\text{NN}$ (solid blue line) and $\hat{\mathcal{L}}_\text{Ryd}^\text{NN}$ (solid green line). This monotonic increase is caused by longer lifetimes at higher $n$, despite the decrease in transport speed caused by lower Rabi frequencies. The scaling of the transport speed with $n$ can be understood by examining the effective interaction strength $J_\text{max}(n)$, whose amplitude can be expressed as the ratio of two terms,
\begin{equation}
        \lvert J_\text{max}(n) \rvert = \frac{\Omega(n)^2 / 4 \lvert \delta_0(n) \rvert}{\lvert \delta_0(n) \rvert/V_\text{max}(n) + 1},
\end{equation}
where $\delta_0(n) =-|\delta_0(n)|< 0$. For small values of $n$, $|\delta_0|\gg V_\text{max}$, such that the $J_\text{max}$ is dominated by the denominator (see Fig.~\ref{fig:principal_quantum_number}b). For larger values of $n$, $|\delta_0|\ll V_\text{max}$, such that the denominator goes to one, resulting in $|J_\text{max}(n)|\sim\Omega^2(n)\sim n^{-3}$. Therefore, there exists a critical value $n_c$ at which $J_\text{max}$ reaches a maximum, after which it decays according to a power law. Despite slower transport at large $n$, the corresponding increase in lifetime, $\tau(n) \sim n^3$, leads to a flattening of the transport probability, reflecting its asymptotic behavior. This plateau is reflected in the ratio between the transport time and the Rydberg state lifetime (see Fig.~\ref{fig:principal_quantum_number}c).

The monotonic increase in transport probability is interrupted by introducing long-range interactions to the Rydberg model, i.e., when considering the transport generated by  $\hat{\mathcal{L}}_\text{Ryd}^\text{LRI}$ (see solid purple line in Fig.~\ref{fig:principal_quantum_number}a). In this case, the transport probability attains its maximum at $n^*=71$ (purple square) for $L=16$. The relative loss in transport probability beyond $n^*$ is dominated by the existence of next-nearest-neighbor (NNN) interactions, which modify the energy spectrum and eigenstates of the chain~\cite{Kay2006}. While the dynamics remains restricted to the $m = 1$ excitation subspace (see Fig.~\ref{fig:principal_quantum_number}d), the NNN interactions generate coherence among distant spins that hinders the constructive refocusing of the spin excitation at the target spin. 

\begin{figure}[t!]
\includegraphics[width=8.6cm]{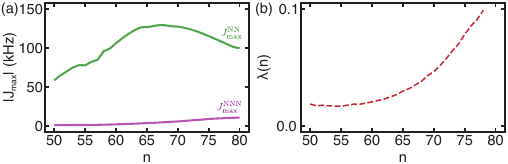}
\caption{
\textbf{Relative contribution of next-nearest-neighbor interactions.}
(a)~Maximum effective interaction strength, $|J_\text{max}(n)|$, between nearest-neighbors (NN, green) and next-nearest-neighbor (NNN, purple) interactions for the Rydberg model. 
(b)~The ratio $\lambda(n) = J_\text{max}^{\mathrm{NNN}} (n) / J_\text{max}^{\mathrm{NN}}(n)$ increases with $n$ due to the growing relative contribution of $J_\text{max}^{\mathrm{NNN}}$ at larger $n$.
}
\label{fig:nnn_effects}
\end{figure}

The loss in transport probability in the LRI model due to NNN interactions can be explained by the competition between the maximum effective interaction strengths $J_\text{max}$ achieved for NN and NNN interactions (see Fig.~\ref{fig:nnn_effects}a). While $J_\text{max}^\text{NN}$ reaches a local maximum at some critical $n_c$ near the point where $V_\text{max}(n)$ overtakes $\delta_0^*(n)$, $J_\text{max}^\text{NNN}$ monotonically increases. Their ratio, which is given by
\begin{align}
    \lambda(n) &= \frac{J_\text{max}^{\mathrm{NNN}}(n)}{J_\text{max}^{\mathrm{NN}}(n)}\\
    &\approx \frac{\lvert \delta_0^{*\text{NN}}(n) \rvert/V^\text{NN}_\text{max}(n) + 1}{\lvert \delta_0^{*\text{LRI}}(n) \rvert / V^\text{NNN}_\text{max}(n) + 1},
\end{align}
is more or less constant until $n_c$, after which it starts diverging (see Fig.~\ref{fig:nnn_effects}b). This divergence reflects the growing influence of NNN interactions relative to NN interactions, which hinders the realization of perfect transport. At the cost of having to solve an inverse eigenvalue problem, this loss in transport probability may be suppressed by including NNN interactions when solving the inversion problem~\cite{Kay2006}, effectively choosing a different solution than the linear spectrum one.

\section{Quantum channel for entanglement distribution}\label{sec:entanglement}
By enabling the transport of single-spin excitations, Rydberg chains can be used as coherent quantum channels to distribute entanglement between distant quantum nodes~\cite{Bose2003, Yang2019}. The entanglement distribution protocol involves adding an auxiliary spin at the beginning of the chain, entangling it with the first spin of the chain, decoupling it from the dynamics by nulling its Rabi frequency, storing its coherence in some long-lived clock state, and then letting the chain propagate the state for $t_\pi$ until it reaches the other end of the chain~\cite{Bose2003, Yang2019}. This protocol results in, e.g, the Bell state $\Phi_+^{0,1}=(\ket{0,0} + \ket{1,1})/\sqrt{2}$ being transported to $\Phi_\phi^{0,L}=(\ket{0,0} + e^{i\phi(L)}\ket{1,L})/\sqrt{2}$, where the length-dependent phase factor can be accounted for by rotating the local measurement basis of the end spin. The entanglement between the auxiliary and end spins can be detected using an entanglement witness~\cite{Peres1996, Horodecki1996, Sun2020}, as well as quantified using an entanglement measure, which we choose as the concurrence~\cite{Wootters1998}, $C^{0,L}_\pi=\max{(0,2p_\pi(L)-1)}$. The concurrence for the single-spin transport protocol depends on the transport probability, and it vanishes when $p_\pi(L)\le1/2$. A nonzero concurrence indicates that the shared state contains distillable entanglement, enabling the implementation of teleportation protocols~\cite{Bennett1993}. 

Because $p_\pi(L)$ decreases with $L$ due to radiative decay over longer transport times, there exists a critical chain length $L_c$, beyond which the Rydberg chain can no longer distribute quantum entanglement. Beyond this cutoff, the two-qubit joint state behaves effectively as a classical mixed state, rendering the quantum channel ineffective. To estimate $L_c$, we numerically compute the set of parameters that achieves the optimal transport probability $p_\pi^*(L)$ for different chain lengths. Because Rydberg state decay depends on temperature, we compute $L_c$ for both ambient conditions ($300~\text{K}$) and cryogenic temperatures ($4~\text{K}$), where coupling to blackbody radiation is strongly suppressed for cryogenic temperatures~\cite{Schymik2021}. 

\begin{figure}[t!]
\includegraphics[width=8.6cm]{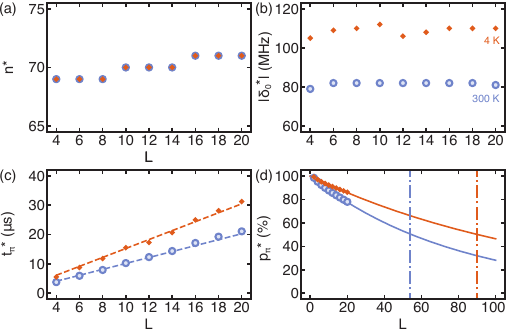}
\caption{
\label{fig:length_scaling}
\textbf{Scaling with chain length.}
(a-b) Optimal principal quantum number $n^*$ and detuning $\delta_0^*$ for the Rydberg model with long-range interactions (LRI) and decay, computed as a function of the chain length $L$ at temperatures of $300~\text{K}$ (blue disks) and $4~\text{K}$ (orange diamonds).
(c)~The optimal transport time increases approximately linearly with $L$.
(d)~The transport probability decreases approximately exponentially (solid line) with $L$, with a smaller decay rate at $4~\text{K}$ due to reduced Rydberg state decay. The maximum chain lengths over which entanglement can be distributed are estimated to be $L_c=54$ and $L_c=90$ at 300 K and 4 K, respectively.
}
\end{figure}

As the chain length is increased, we find that the optimal principal quantum number $n^*$ has a slowly increasing piecewise-continuous pattern (see Fig.~\ref{fig:length_scaling}a), while the optimal detuning values remain relatively constant (see Fig.~\ref{fig:length_scaling}b). The stability of the control parameters results in a relatively stable $J_\text{max}(n)$, such that the transport time grows approximately linearly with $L$ for chains up to our largest simulated chain length $L=20$ (see Fig.~\ref{fig:length_scaling}c). This increase in transport time results in a decrease in $p_\pi^*(L)\sim\exp{(-t_\pi^*(L) / \tau(n^*))}$ (see Fig.~\ref{fig:length_scaling}d), which decays approximately exponentially.
By extrapolating our simulation data to larger chain lengths, we find that the maximum chain length for distributing entanglement is $L_c = 54$ at $300~\text{K}$ and $L_c = 90$ at $4~\text{K}$. Reducing the temperature thus provides an additional practical benefit beyond increasing vacuum lifetime and Rydberg decay rates. 


\section{Conclusion}\label{sec:conclusion}
In conclusion, we have shown that Rydberg chains can act as a coherent quantum channel to distribute quantum information at the quantum speed limit between distant atoms separated by hundreds of microns. These chains provide an alternative approach to increasing the connectivity of quantum processors, complementing existing methods based on mediating quantum gates, shared bosonic modes, photonic links, and the displacement of quantum particles.

Through numerical simulations, we have shown that finding the set of optimal control parameters for realizing the perfect transport condition in Rydberg chains requires balancing competing trade-offs imposed by approximation errors, physical errors, and control limitations. These trade-offs set an upper bound on the ultimate performance achievable by the spin chain as a coherent quantum channel. In practical settings, this performance will be further degraded from spatial inhomogeneities and temporal instabilities in both control and physical parameters, e.g., resulting from spatial disorder and motional dephasing. This might be quantified using the transport probability, providing a figure of merit to quantify detrimental effects and optimize the performance of robust control protocols.

While our approach relies on mapping the native Rydberg spin model onto a Heisenberg~XX model using second-order perturbation theory, our analysis framework can readily be applied to quantify the performance of alternative protocols based on resonant dipole-dipole interactions~\cite{Gunter2013, Schonleber2015, Barredo2015, Orioli2018}, exchange between Rydberg and Rydberg-dressed ground states~\cite{Schempp2015, Letscher2018}, time-dependent driving~\cite{Wang2025}, and second- and fourth-order exchange of ground hyperfine states coupled to Rydberg states~\cite{Wuster2011, Glaetzle2015, vanBijnen2015, Li2022}. It might also be extended to topological quantum phenomena capable of distributing quantum information, such as anomalous chiral edge flow in the Kitaev honeycomb model~\cite{Po2017, Nishad2023}. Beyond its experimental realization in Rydberg chains, this work could be extended to arbitrary spin networks with multiple Rydberg excitations to explore transport and scattering phenomena, as well as applications such as quantum state routing.

\section{Acknowledgment}
This research was funded thanks in part to the Canada First Research Excellence Fund (CFREF). We thank Fan Yang, Christopher Wyenberg, and Paola Cappellaro for fruitful discussions. We acknowledge early contributions by Pavan Jayasinha.

\providecommand{\noopsort}[1]{}\providecommand{\singleletter}[1]{#1}%

\end{document}